\documentclass[conference]{IEEEtran}
\IEEEoverridecommandlockouts
\usepackage{cite}
\usepackage{amsmath,amssymb,amsfonts}
\usepackage{algorithmic}
\usepackage{graphicx}
\usepackage{textcomp}
\usepackage{xcolor}
\usepackage{hyperref}
\def\BibTeX{{\rm B\kern-.05em{\sc i\kern-.025em b}\kern-.08em
    T\kern-.1667em\lower.7ex\hbox{E}\kern-.125emX}}

\begin{document} 

\title{Shot Allocation Strategies \\ in Debugging Quantum Circuits}
%{\Large Era Thaqi and Margaret Martonosi\\Princeton University
%{era.thaqi@princeton.edu}}
\author{
\IEEEauthorblockN{Era Thaqi, Coco Gong, Margaret Martonosi} \IEEEauthorblockN{
%\textit{Department of Computer Science} \\ 
\textit{Princeton University }
%\\ mrm@princeton.edu} 
} }
\maketitle

\begin{abstract}
Debugging quantum programs remains challenging due to the nature of quantum measurement. As quantum algorithms and systems scale, it is increasingly crucial to develop rigorous debugging and verification approaches.  Statistical assertion-based debugging validates quantum program behavior through repeated circuit execution and statistical hypothesis testing, but existing frameworks treat shot count as a fixed input parameter, leaving the problem of optimal shot allocation across multiple mid-circuit assertions unexplored. To improve debugging efficiency by making better use of given shot counts, this paper proposes and evaluates four shot allocation strategies within the QDB quantum debugger implemented within the CUDA-Q framework.  These allocation approaches---equal split, static weight allocation, adaptive entropy routing, and sequential retirement---use statistical approaches to test whether qubits have expected state characteristics at different points in execution. We evaluate these allocation strategies on different circuits, under both noise-free and depolarizing noise conditions. We show how adaptive strategies use shots efficiently by concentrating them on assertions with high uncertainty. This substantially reduces total shot usage while maintaining comparable debugging confidence.
\end{abstract}


\section{Introduction}
Quantum programs are difficult to debug;  intermediate quantum states cannot be inspected directly without collapsing the state. Statistical assertions address this by using repeated measurements allowing developers to test whether a qubit is in a classical, uniform superposition, or entangled state\cite{yipengpaper,ncstatedynamicdebug}. 

Previous work implemented the QDB assertion-tool in both IBM Qiskit \cite{yipengpaper} and CUDA-Q \cite{li2025cudaq_assertions} environments. Either in state vector simulations or in runs on prototype hardware, assertions are evaluated through repeated calls to \texttt{cudaq.sample}, and the number of shots determines how many independent executions are used to form the measurement distribution \cite{li2025cudaq_assertions}. However, these workflows treat the shot count as a fixed input parameter and do not address how a given overall shot budget should best be distributed across multiple assertion checkpoints. This limitation is important because the number of shots assigned to each checkpoint directly affects both the statistical reliability of the assertion and the overall cost of debugging.  

Because quantum assertions infer state properties from sampled measurements, insufficient shots can cause statistical assertions to fail even in the absence of real bugs, creating a sampling hazard under finite hardware budgets. In this paper, we experimentally evaluate four shot allocation strategies within QDB's CUDA-Q assertion framework: equal split, static weighting, adaptive entropy routing, and sequential retirement. Across a set of benchmarks, under both noiseless and depolarizing-noise settings, we compare these strategies using wasted-shot counts, false positive rates, and bug-detection power.

Our results show that assertion-based quantum debugging becomes substantially more effective when shots are allocated adaptively across checkpoints, rather than simply dividing shots equally among the various assertions in the code. For example, assertions that a qubit is in a ``classical'' state typically converge with fewer shots than assertions regarding product states or uniform distributions.  As a result, rebudgeting more shots toward statistically-uncertain checkpoints improves bug detection power by up to 13\%, while reducing wasted sampling. Under noise, adaptive strategies improve statistical reliability, with Strategy 3 achieving the lowest false-positive rates and the largest reduction in wasted shots.

%\subsection{Statistical Assertion Framework}
%In this paper, we used three types of assertions: a classical, uniform and product-state assertion. 
%A classical assertion checks whether the observed outcomes are consistent with a single deterministic expected bitstring. A uniform assertion is a special case of a superposition assertion and checks whether the observed outcomes are consistent with a uniform distribution over the computational basis states of the measured register. A product assertion checks whether two measured states behave independently \cite{product_state}. Each assertion is evaluated from a finite sample of measurement outcomes, so correctness is not observed directly but inferred statistically. A low p-value ($p < 0.05$) indicates that the observed data are unlikely under the asserted behavior and thus suggests a violation of the assertion \cite{li2025cudaq_assertions}. Each assertion returns a p-value indicating whether the sampled outcomes are statistically consistent with the expected behavior.

\section{Comparing Shot Allocation Strategies}

Let $B$ denote the total shot budget for a circuit's execution. Let $A = \{a_1, \dots, a_m\}$ denote the set of assertion checkpoints placed within the circuit's execution. 

\textbf{$S_0$: Equal split.} $S_0$ splits the budget uniformly, assigning each of $m$ assertions $\frac{B}{m}$ shots.

\textbf{$S_1$: Static weights.} $S_1$ uses fixed type-based weights: $10\%$ to classical assertions, $40\%$ to uniform assertions, and $50\%$ to product-state assertions to reflect empirically-observed trends in sampling difficulty. 

\textbf{$S_2$: Adaptive entropy.} This operates in two phases: an initial probing phase allocates a small fraction ($10\%$) of the total budget equally across all assertions to gather preliminary measurement data, after which Shannon entropy is computed for each assertion’s observed distribution. The remaining budget is then allocated proportionally to these entropy values. 

\textbf{$S_3$: Sequential retirement.} Here, shots are assigned in small fixed increments to all active assertions. After these runs, the assertions are evaluated. Checkpoints that have reached a statistically-decisive outcome are removed from consideration, with more shots allocated to remaining checkpoints. % The remaining shot budget is then redistributed among the unresolved assertions. The process continues until all assertions are resolved or the available budget is exhausted. 

\subsection{Entropy Based Scoring}
$S_2$ budgets phase-2 shots based on how far each assertion's empirical distribution is from its expected ideal at the probe phase. We quantify this gap using Shannon entropy. For an empirical distribution $p$ with measured counts over outcomes, Shannon entropy is defined as $H(p) = -\sum_{x} p(x) \log_2 p(x)$, where the sum runs over observed outcomes and $p(x)$ is the empirical probability of outcome $x$. %\cite{shannonentropy}.

\textbf{A classical assertion} claims that the state should be concentrated on a single outcome. This gives target entropy $H_\mathrm{target} = 0$. The uncertainty score is $s_i^\mathrm{classical} = H(p_i)$, where $p_i$ is the empirical distribution from the probe. Values near zero indicate that the observed distribution is concentrated on a single outcome.

\textbf{A uniform assertion} claims that the state is in equal superposition across all $2^n$ basis states of an $n$-qubit register. The target distribution is uniform with target entropy $H_\mathrm{target} = n$. The uncertainty score is $s_i^\mathrm{uniform} = \left| H(p_i) - n \right|$, where $n$ is the number of qubits in the register being checked. Larger values indicate the empirical distribution is concentrated on fewer outcomes than uniform sampling would produce. 

\textbf{A product assertion} claims that two registers $A$ and $B$ should be statistically independent, with joint distribution factors as $p_{AB}(a, b) = p_A(a) \cdot p_B(b)$. 
%Under independence, joint entropy equals the sum of marginal $H(p_{AB}) = H(p_A) + H(p_B)$ \cite{wiki_joint_H}. 
We compute the empirical marginals $p_A$ and $p_B$ by summing joint counts over the other register, then measure the deviation from independence as $s_i^\mathrm{product} = \left| H(p_{AB}) - H(p_A) - H(p_B) \right|$.  Higher entropy scores 
indicate the empirical distribution is more concentrated than the asserted structure predicts.  If above a provided threshold, the assertion fails indicating the state at the 
checkpoint does not sufficiently match the expected behavior.
%However, before computing the entropy score, $S_2$ first checks whether the probe p-value falls in a numerically extreme range (above 0.9999 or below 0.0001). The uncertainty score is set to zero for such assertions, redirecting phase-2 shots to assertions where additional sampling could meaningfully reduce uncertainty.
%

\begin{figure}[t]
    \centering
    \includegraphics[width=0.8\linewidth]{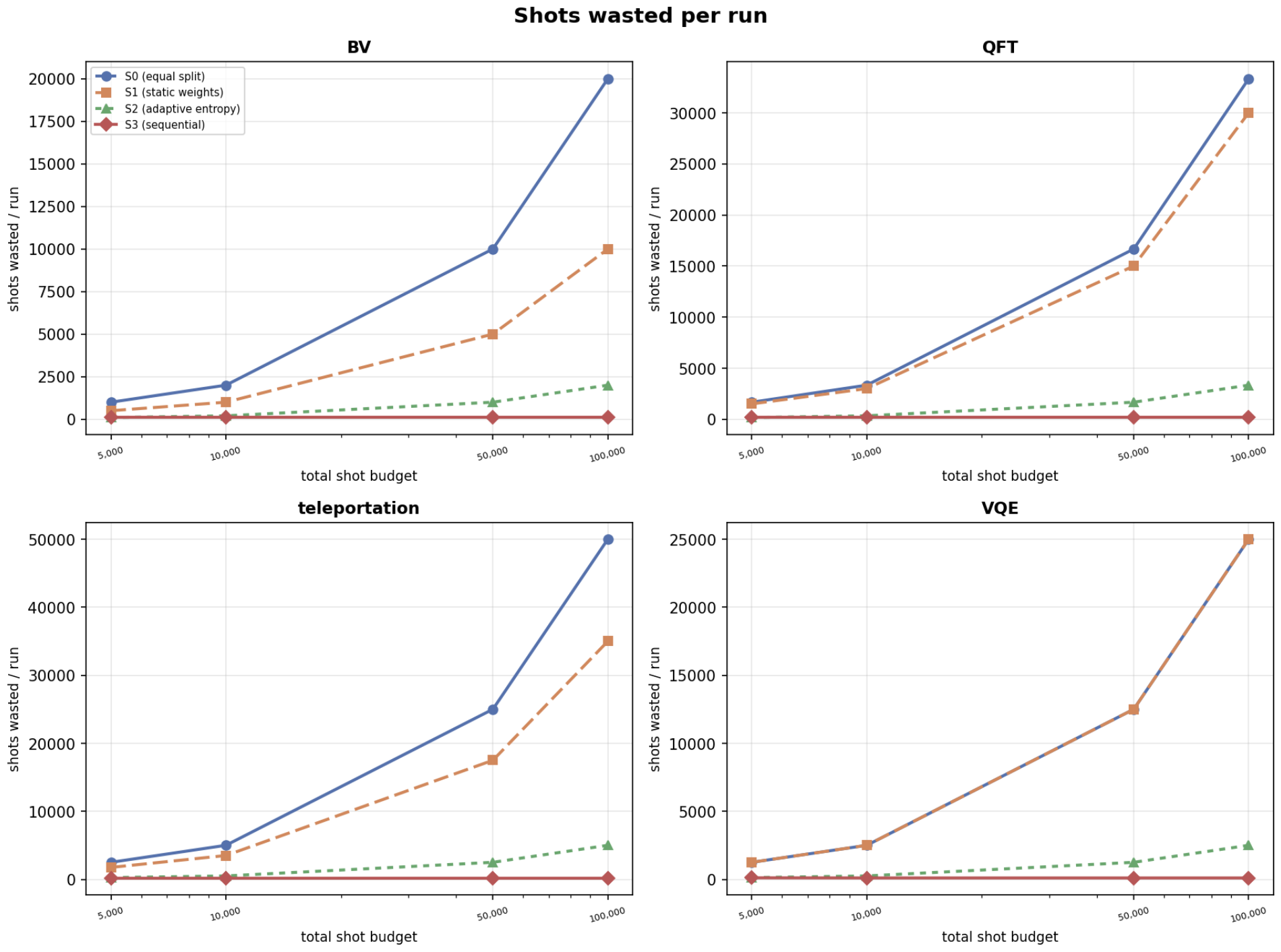}
    \caption{Results across four benchmarks}
    \label{plot1}
\end{figure}

\section{Results and Discussion}
Figure~\ref{plot1} summarizes our results for four benchmarks. Full code and experimental data are available in our GitHub repository:
\href{https://github.com/Era-Thaqi/Shot_optimization}{Shot\_optimization} \cite{shot_optimization_repo}.

{\em Shot utilization and wasted shots:} 
We define wasted shots as shots allocated to assertions beyond the point at which the assertion becomes statistically decisive. 
%In our benchmarks, these include the final classical assertions in BV, the initialization classical and product assertions in QFT, the product-state in teleportation, and the pre-entanglement product-state assertion in VQE.

Across all benchmarks, $S_0$ wastes the most shots, with waste increasing sharply as the total budget grows. 
%At a budget of 100{,}000, it wastes roughly 20{,}000 shots in BV , 30{,}000 in QFT, 50{,}000 in Teleportaton, and 25{,}000 in VQE, reflecting the fraction of decisive assertions in each circuit. 
$S_1$ reduces waste by assigning a smaller fixed fraction of the budget to classical assertions, but still behaves similarly to $S_0$. In contrast, $S_2$ and $S_3$ substantially improve shot efficiency by reallocating shots away from low-uncertainty checkpoints. $S_3$ achieves the best shot efficiency overall. Its retirement approach stops shot allocation once assertions are statistically decisive. 

%\begin{figure}[!t]
   % \centering
    %\includegraphics[width=\columnwidth]{plot1.png}
   % \caption{Shots wasted across strategies.}
    %\label{fig:shot_waste_all}
%\end{figure}

{\em Exploration of assertions that become decisive and why:}
Assertions naturally separate into three regimes based on their p-value behavior across 50 runs at 100 shots per run. %In a well calibrated test under a true null hypothesis, p-values behave approximately as i.i.d samples from a Uniform[0,1] distribution. 
In our experiments, a subset of assertions is immediately decisive: they produce p-values pinned at 0 or 1 with zero variance across runs. The p = 0 cases correspond to strong violations of the asserted property (e.g., structural mismatch such as non-classical states being tested as classical or entanglement detected via Fisher-style independence tests), while the p = 1 cases arise when the observed outcomes exactly match the asserted distribution, leaving no statistical ambiguity. 

The remaining assertions exhibit expected stochastic behavior under the null hypothesis, with p-values spanning the full unit interval and averaging around 0.5, consistent with a Uniform[0,1] distribution. These include uniformity checks and separability/product-state tests. 
%This behaviour is not a result of any particular algorithm structure, but a structural property of the benchmark assertions themselves, determined by the algebraic relationship between each circuit’s state and the property each assertion check

{\em Bug Detection Power vs Shots Allocated:}
Bug detection power is defined as the fraction of repeated trials in which the assertion successfully detected a circuit bug at a given shot count. Here, we insert bugs in the benchmarks to show the assertion's ability to detect them.

We first consider a \emph{challenging} assertion, \texttt{after\_init/uniform} under a subtle initialization bug introduced via a small \(R_y(0.024)\) rotation. %as seen in figure \ref{plot2}. 
Because this bug only slightly biases the output distribution, distinguishing the buggy and correct circuits requires many samples; detection power increases with shot counts. 
In contrast, consider a \emph{decisive} assertion, \texttt{after\_measure/classical}, under a structural bug in which BV's final Hadamard layer is accidentally omitted. This destroys the deterministic output and is detected almost immediately, causing saturation at 1.0 even at very low shot count. 

%\begin{figure}[h]
   % \centering
    %\includegraphics[width=0.8\linewidth]{bugdetection.png}
   % \caption{Results across four benchmarks}
    %\label{plot2}
%\end{figure}

%The markers indicate how different allocation strategies distributed a fixed budget of 50{,}000 shots. For the stochastic assertion ($S_0$, $S_1$) each allocate 10{,}000 shots, yielding power of roughly 0.75. In contrast $S_2$ allocated 13{,}107 shots and $S_3$ allocated 12{,}415, increasing detection power to 0.85. Since the classical assertion is already saturated after very few shots, the additional budget assigned is essentially wasted. 

%\subsection{False Positive Rates}
\section{Conclusion}
%We studied how a fixed shot budget should be distributed across statistical assertions in quantum debugging using the CUDA-Q assertion framework. 
Adaptive strategies for shot distribution, particularly entropy-based routing and sequential retirement, reduce wasted sampling on structurally decisive checkpoints and redirect that budget toward stochastic assertions, which yields higher bug-detection power and lower false-positive rates under noise at the same total cost.
\bibliographystyle{IEEEtran}
\bibliography{refs}

@inproceedings{yipengpaper,
author = {Huang, Yipeng and Martonosi, Margaret},
title = {Statistical assertions for validating patterns and finding bugs in quantum programs},
year = {2019},
isbn = {9781450366694},
url = {https://doi.org/10.1145/3307650.3322213},
doi = {10.1145/3307650.3322213},
booktitle = {Proc. 46th Intl. ACM/IEEE Symposium on Computer Architecture},
location = {Phoenix, Arizona},
series = {ISCA '19}
}

@inproceedings{li2025cudaq_assertions,
  author    = {Li, Jocelyn and Rubinshtein, Ella and Martonosi, Margaret},
  title     = {Statistical Assertions for Debugging Quantum Circuits and States in CUDA-Q},
  booktitle = {2025 IEEE International Conference on Quantum Computing and Engineering (QCE)},
  year      = {2025},
  pages     = {2074--2081},
  publisher = {IEEE}
}

@inproceedings{ncstatedynamicdebug,

  author    = {Ji Liu and Gregory T. Byrd and Huiyang Zhou},

  title     = {Quantum Circuits for Dynamic Runtime Assertions in Quantum Computation},

  booktitle = {Proc. 25th Intl. Conf. on Architectural Support for Programming Languages and Operating Systems (ACM ASPLOS '20)},

  year       = {2020},
  doi        = {10.1145/3373376.3378488},

  url        = {https://doi.org/10.1145/3373376.3378488}

}

@misc{shot_optimization_repo,
  author       = {Thaqi, Era},
  title        = {Shot Optimization},
  year         = {2026},
  publisher    = {GitHub},
  howpublished = {\url{https://github.com/Era-Thaqi/Shot_optimization}},
  note         = {Accessed: 2026-06-13}
}
\end{document}